\documentclass[final,5p,times,twocolumn]{elsarticle}

\usepackage{amsmath,amssymb,stmaryrd, mathtools}

\usepackage{graphicx}
\usepackage{enumitem}
\usepackage{float}
\DeclareUnicodeCharacter{2212}{-}
\usepackage[caption=false]{subfig}
\usepackage[font=small,justification=centering,labelfont=bf]{caption}
\usepackage{url}
\usepackage{breakurl}
\usepackage[T1]{fontenc}
\usepackage[utf8]{inputenc}
\usepackage{graphicx} 
\usepackage{cancel}
\usepackage{xcolor}
\usepackage[table]{xcolor}
\usepackage{array}
\usepackage{float}

\journal{Computer Physics Communications}

\begin{document}

\begin{frontmatter}



\title{\textit{fair\_data.py}: implementing FAIR data compliance in Tribchem}


\author[inst1]{Lucrezia Berghenti}
\author[inst1]{Elisa Damiani}
\author[inst1]{Margherita Marsili}
\author[inst1]{Maria Clelia Righi}

\affiliation[inst1]{organization={Department of Physics and Astronomy, University of Bologna},
            addressline={Viale Carlo Berti Pichat 6/2}, 
            city={Bologna},
            postcode={40127}, 
            state={Italy}}

\begin{abstract}
The increasing complexity and volume of data generated by high-throughput computational materials science require robust tools to ensure their accessibility, reproducibility, and reuse. In particular, integrating the FAIR Guiding Principles (Findable, Accessible, Interoperable, and Reusable) into computational workflows is essential to enable open science practices. TribChem is an open-source Python software developed for the automated simulation of solid-solid interfaces using density functional theory (DFT). While TribChem already incorporates several FAIR-aligned features, we present here a dedicated FAIR utility designed to transform TribChem results into FAIR-compliant datasets. This utility comprises two tools: \textit{fair\_data.py}, which automatically generates standardized machine- and human-readable outputs from the TribChem database, and \textit{retrieve\_data.py}, which facilitates efficient data extraction through a keyword-based interface. In this paper we show the capabilities of the fair utility with examples for bulk, surface, and interface systems. The implementation allows seamless integration with public repositories such as Zenodo, paving the way for reproducible research and fostering data-driven materials discovery.

\textbf{Program summary}

\textit{Program Title:} fair\_data.py, retrieve\_data.py

\textit{Developer’s repository link: }https://gitlab.com/triboteam/fair-data.git

\textit{Programming language:} Python

\textit{External libraries:} pymongo\cite{mongodb}, bson\cite{bson}

\textit{Licensing provisions:} Creative Commons Attribution 4.0 International (CC BY 4.0)

\textit{Nature of problem:} High-throughput first-principles simulations produce large amounts of data, but these results are often stored in formats that are not easily shareable, interoperable, or reusable. In particular, while the TribChem code stores structured data in a MongoDB database, there is no straightforward method to export the results in a format that complies with the FAIR data principles, limiting accessibility and long-term reuse.

\textit{Solution method:} We developed two Python command-line tools to integrate FAIR data practices into the TribChem workflow. The \textit{fair\_data.py} script connects to the MongoDB database used by TribChem, retrieves the relevant document based on user-defined parameters (such as system type, material ID, and Miller indices), and generates both machine-readable (JSON) and human-readable (TXT) files containing computational results and metadata. The \textit{retrieve\_data.py} script enables selective data extraction from the JSON output by keyword-based queries, allowing efficient reuse and analysis of FAIR-formatted data.

\end{abstract}




\end{frontmatter}


\section{Introduction}
\label{cha:Introduction}
The rapid advancement of computational materials science has led to an unprecedented generation of scientific data, especially in the high-throughput computational frameworks. However, the scientific value of data is often limited by challenges in accessibility, interoperability, and long-term preservation.
In response to these challenges, the scientific community has embraced the FAIR Guiding Principles (Findable, Accessible, Interoperable, and Reusable), which provide a framework for ensuring that research data can be effectively discovered, accessed, integrated, and reused by both humans and machines. Originally formulated in 2016, these principles have become increasingly critical as computational workflows generate ever larger datasets with greater complexity and diversity.
TribChem, open-source software designed by our group for the automatized first-principles simulation of solid-solid interfaces, exemplifies the modern computational approach to materials research. Built upon established workflow managers like Atomate and Fireworks, and utilizing MongoDB for data storage, TribChem enables systematic investigations of bulk materials, surfaces, and interfaces through automated DFT calculations. The software's modular architecture facilitates the calculation of various properties, from basic structural parameters to complex interfacial phenomena such as adhesion energies and charge redistribution.
While TribChem's existing infrastructure incorporates several elements that align with FAIR principles, such as structured database storage, standardized data formats through Pymatgen integration, and connectivity with the Materials Project database, there remained significant opportunities for enhancement in providing users with straightforward tools to export and share their computational results in a FAIR-compliant manner.
To address this need, we have developed a dedicated FAIR utility for TribChem that seamlessly integrates FAIR compliance into our existing computational workflows. This utility consists of two complementary tools: \texttt{fair\_data.py}, which automatically generates FAIR-compliant outputs from TribChem results, and \texttt{retrieve\_data.py}, which facilitates efficient data extraction and analysis. By implementing these tools, we aim to bridge the gap between our automatized high-throughput computational data generation and the principles of open, reproducible science.

\section{Background}
\label{cha:Background}

Tribchem is a software designed to perform the high-throughput study of solid interfaces. The software is composed of three main units for the study of bulks, surfaces, and interfaces \cite{Tribchem}\cite{pub_82}. Its modular structure allows for the execution of different types of calculations and can be easily extended to include new features. The program is based on Atomate\cite{atomate} and Fireworks\cite{fireworks} workflow manager and relies on MongoDB\cite{mongodb} for database communication.
Tribchem is entirely written in Python and uses different packages from Materials Project\cite{Materials_Project}. To perform DFT calculations it relies on the Vienna Ab initio Simulation Package (VASP) \cite{vasp_code}\cite{vasp_algo}\cite{PAW_method}\cite{PAW_implementation}.
The Table \ref{table_quantities} schematically shows the different properties that Tribchem allows to calculate.
From a worklow perspective, Tribchem is designed in 3 main stages and each one is implemented in TribChem using Python code structured into separate modules. Namely these are: the bulk module, the surface module and the interface module. 

The bulk module generates and relaxes the bulk structure to ensure that the lattice parameters and atomic positions are optimized. This is the starting point for the generation of the surface structure. In order to do that, the bulk modulus and the volume are converged with respect to the cutoff energy. Calculations are performed for increasing values of the energy cutoff and the bulk modulus and cell volume are extrapolated fitting the data with Birch-Murnaghan\cite{Murnaghan} equation of state. The convergence is reached when the difference between two
 consecutive steps is below $1\%$ for both bulk modulus and volume. The command that executes the bulk workflow is the following: \textit{tribchem workflow converge\_bulk mids="[mp-8062]" formulas="SiC"}.
With the same tribchem command, it is also possible to identify the optimal k-point sampling. In this case, TribChem fixes the energy cutoff at the optimal value while varying the k-point density until convergence is reached.

The surface module creates surface slab from the optimized bulk, given the specified Miller index provided by the user. The optimal thickness is calculated by converging the surface energy $E_\gamma$: the slabs total energies $E_\text{slab}(N)$ at different layer thicknesses $N$ is fitted using the following formula, obtained by inverting the surface energy definition\cite{Tribchem}:
\begin{equation}
    E_\text{slab} = E_\gamma * A + N * N_\text{at}*\epsilon_\text{bulk}
\end{equation}
where $N_\text{at}$ is the number of atoms per layer, $\epsilon_\text{bulk}$ is the bulk cohesive energy, $A$ is the slab in-plane area, and $N$ is the number of layers in the slab system.
The surface energy $E_\gamma$ is therefore a parameter obtained by the fit when convergence is reached.
The command to execute this workflow is the following: \textit{tribchem workflow converge\_slab mids="[mp-150]" formulas="Fe" miller="[[1, 1, 1]]"}.

It is also possible to insert a custom slab in the database without starting from a stored bulk. In this case the command to launch is the following: \textit{tribchem workflow converge\_slab mids="[mp-8062]" formulas="SiC" miller="[[1, 1, 1]]" input\_dir="my\_path/folder" functional="PBE" -cs="PBE.custom\_slab"}. The user has to specify the path of the directory (\textit{input\_dir}) where the input files are contained, the functional used (\textit{functional}) and the name of the collection (\textit{cs}) where the data should be stored in the database.

The first step of the interface module is to build the interface by matching the two slabs. The surface matching is performed by the Pymatgen library\cite{pymatgen}, based on the Zur algorithm\cite{Zur}, which builds the lowest area supercell meeting a series of criteria (on the maximum allowed lattice sides and angles strains, and supercell areas). 
Once the interface between two surfaces is constructed, the interfacial adhesion energy is evaluated by computing the interaction energy for different relative lateral displacements. These displacements correspond to shifts that align high-symmetry points of the two mating surfaces. This procedure helps identify the most stable configuration which is the one with the minimum interaction energy.

The number of non-equivalent lateral positions depends on the nature of the interface: for homogeneous interfaces (same material on both sides), there are typically 6 distinct lateral configurations\cite{pub_49}\cite{pub_55}; while for heterogeneous interfaces (different materials), an increasing number of configurations may be considered due to the lower symmetry\cite{pub_70}\cite{pub_82}\cite{pub_101}.

Once the interface is optimized, it is possible to calculate several interface-related quantities: the potential energy surface (PES)\cite{pub_23}, the adhesion energy, the perpendicular potential energy surface (PPES), the charge displacement\cite{pub_50} and the shear strength.

The potential energy surface (PES) provides a map of the interaction energy as a function of relative lateral in-plane displacements between the two surfaces, allowing for the identification of energetically favorable stacking configurations. 
By definition, the adhesion energy $E_\text{adh}$ is the difference in energy between the interface system $E^{12}_\text{interface}$ in its most stable configuration and the two isolated slabs, $E^{1}_\text{slab}$ and $E^{2}_\text{slab}$:
\begin{equation}
    E_\text{adh}=\dfrac{1}{A}[E^{12}_\text{interface}-(E^{1}_\text{slab}+E^{2}_\text{slab})].
\end{equation}
Therefore, the adhesion energy quantifies the energetic gain upon interface formation and is a key descriptor of interfacial stability.

The perpendicular potential energy surface (PPES) extends this concept along the direction normal to the interface, capturing the variation of interaction energy with respect to the interlayer distance. This is particularly relevant for studying interface separation, exfoliation processes, and van der Waals interactions\cite{PPES_graphites}\cite{PPES_exfoliation}\cite{PPES_VdW}.

The charge displacement, computed as the difference between the charge density of the full interface and the sum of the charge densities of the isolated slabs, provides insight into the electronic reorganization induced by interface formation. This quantity is essential to evaluate the presence of charge transfer, dipole formation, or interfacial polarization effects\cite{pub_50}, \cite{pub_82}, \cite{pub_96}.

\begin{table*}[htbp!]
\centering
\begin{tabular}{|c|l|}
\hline
\rowcolor{yellow!20}
\textbf{System Type} & \multicolumn{1}{c|}{\textbf{Computed Quantities}} \\ \hline
\textbf{Bulk} & 
\begin{tabular}[t]{@{}l@{}}
- Optimal plane-wave cutoff energy \\
- k-point density \\
- Lattice parameters \\
- Bulk modulus \\
- Cohesive energy
\end{tabular} \\ \hline
\textbf{Surfaces} &
\begin{tabular}[t]{@{}l@{}}
- Optimal slab thickness \\
- Surface energy
\end{tabular} \\ \hline
\textbf{Interfaces} & 
\begin{tabular}[t]{@{}l@{}}
- Automated interface generation (Zur algorithm) \\
- Potential energy surface (PES) \\
- Adhesion energy \\
- Perpendicular Potential Energy Surface (PPES) \\
- Charge redistribution at the interface
\end{tabular} \\ \hline
\end{tabular}
\caption{The table shows the physical quantities that TribChem computes for various system types, including bulk materials, surfaces and interfaces properties.}
\centering
\label{table_quantities}
\end{table*}


When performing a high-throughput study, the amount of data generated is typically high; therefore, it becomes necessary to manage them in an efficient way. 
During Tribchem installation, the user creates several databases.
The most relevant databases are:
\begin{itemize}
    \item \texttt{Fireworks database}, which contains the data relative to the workflow execution, the simulation results and some relevant metadata (such as the location of the VASP output files). Indeed, to store workflows, the Fireworks library uses MongoDB\cite{mongodb}, which is a NoSQL database that uses JSON-like documents to store and manipulate data.
    \item \texttt{Tribchem database}, which is a custom high-level database to save results and relevant information of the high-throughput calculations facilitating their retrieve and sharing.
\end{itemize}
Among these, in the FAIR data perspective, the Tribchem database plays a central role since it collects the key outcomes of the calculations and represents the primary resource queried by users during data analysis and post-processing. This database includes several collections divided into the different kind of structures and functionals used for performing the DFT calculations\cite{Tribchem}. For example, the main collections where the data related to bulk, surface and interface structures, simulated at the DFT-PBE level, are stored are respectively the \texttt{PBE.bulk\_elements}, \texttt{PBE.slab\_elements} and \texttt{PBE.interface\_elements}.

Each element is identified by a set of identifiers that depends on the collection where the data is stored:
\begin{itemize}
    \item the bulk elements are labelled by the \texttt{mid}, which is the material identifier usually corresponding to the Materials Project ID for a given material (it could also be an alphanumeric value); in addition it is also possible to use the identifiers \texttt{formula} and \texttt{name} corresponding to the chemical formula and the name of the material (the latter has been introduced to avoid misinterpretation for material having the same chemical composition)
    \item the slab elements, in addition to the bulk identifiers, have also the \texttt{miller} identifier which is a Python list in the form $[h, k, l]$ that defines the crystalline orientation;
\item the interface elements have the identifiers obtained by combining the identifiers of the two slabs forming the interface, meaning \texttt{mid=mid1\_mid2}, \texttt{formula=f1\_f2}, \texttt{miller=[h1,k1,l1]\_[h2,k2,l2]}.
\end{itemize}

More detailed information can be found in the Tribchem user guide: https://triboteam.gitlab.io/tribchem/.

The high-throughput module is implemented making use of the FireWorks library.
Fireworks defines every workflow using three hierarchical components: the FireTask (basic unit of work), the FireWork (a group of FireTasks), and the workflow (a network of FireWorks). The workflow is the only unit that can be launched by the user on a workstation. Tribchem includes the implementation of several workflows for bulks, surfaces and interfaces. 
The bulk workflow, relying on the corresponding bulk modulus, consists in the following steps: the structure is retrieved from Materials Project Database, geometry optimization (relax of the structure), computation of the total energy and storing of the relaxed structure and computed data in the database.
The workflows relative to the slab create an optimize surface from bulk materials. To do that, first convergence tests are performed starting from stored bulk structures to determine optimal slab thickness and vacuum spacing. Otherwise the user can insert directly a surface structure into the database using the 'Custom slab' insertion option. Then the surface geometry is optimized and the surface properties are calculated. A workflow that allows to construct monolayered 2D material, from layered materials bulk has also been developed.
Starting from the slabs, there is a workflow that matches the two slabs to create the interface. From this it is possible to calculate several interfacial properties such as the adhesion energy, the potential energy surface (PES), the Perpendicular Potential Energy Surface (PPES) and the Charge redistribution at the interface.

TribChem, as an automatized, high-throughput computational infrastructure for interface calculations, generates massive amounts of computational data.
In general, in the context of modern computational science, data production and management is as much central as models and algorithms. To ensure that scientific data can be effectively reused, shared and preserved, the FAIR Guiding Principles were introduced, where the term 'FAIR' stands for: Findable, Accessible, Interoperable and Reusable.
First formally published in 2016 \cite{FAIR_guidelines}, the FAIR principles aim to improve the ability of both humans and machines to discover, access, integrate, and reuse data. They are especially relevant in an era of increasing data volume, complexity, and production speed in scientific research.
TribChem already partially incorporates elements that align with FAIR principles including: a MongoDB high-level database for structured data storage, Pymatgen integration for standardized structure manipulation, and Materials Project connectivity for interoperability, nevertheless implementing a dedicated FAIR utility would enhance its compliance with FAIR data principle significantly.  The benefits include enabling scientific reproducibility, fostering community collaboration in tribology and materials research, and facilitating data reuse for machine learning and materials discovery. It would also ensure long-term data preservation and connect with other major materials databases like NOMAD \cite{NOMAD} or Zenodo\cite{Zenodo}.
For this reasons, we decided to implement the Fair utility which is described in the next section. 

\section{Implementation and examples}
\label{cha:Implementation_and_examples}

Within computational workflows, such as those based on Tribchem, FAIR principles can be integrated directly at the point of data generation.

The process involves three key steps:
\begin{enumerate}
    \item Run Tribchem simulations;
    \item Generate FAIR outputs using \texttt{fair data.py};
    \item Upload data and metadata to a public repository.
\end{enumerate}

\begin{figure*} [htbp!]
    \centering
    \includegraphics[width=1.0\linewidth]{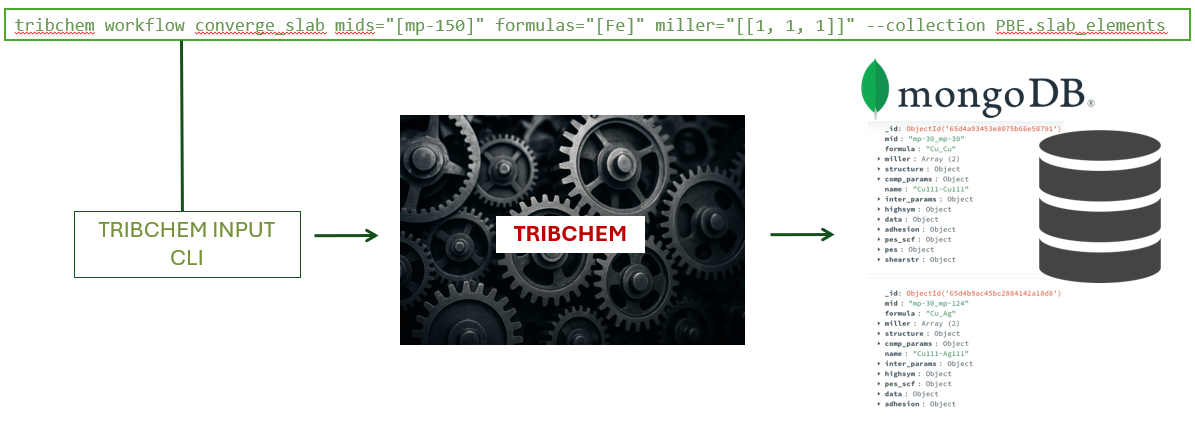}
    \caption{Schematic representation of a Tribchem workflow (e.g. surface energy convergence): the user provides input via CLI, Tribchem executes the calculations and outputs the results which are stored in a specific MongoDB collection (e.g. PBE.slab\_elements).}
    \label{fig:tribchem_pipeline}
\end{figure*}

The first step is schematically represented in Fig. \ref{fig:tribchem_pipeline} which shows the execution of the standard Tribchem workflow.
The user provides the input via CLI, Tribchem executes the calculations and the results are stored in the collection specified in the input command.

The second step involves the specific utility that we have created to conform the data produced by Tribchem to the FAIR principles. 
The script connects to MongoDB database and retrieves the document relative to the system specified in the input command.

The output files generated by \texttt{fair data.py} can be uploaded into a public repository, such as Zenodo. This ensures that each dataset is enriched with the necessary metadata and formatted for long-term reuse.

\subsection{\texttt{fair\_data.py} script}
\begin{figure*} [htbp!]
    \centering
    \includegraphics[width=1.0\linewidth]{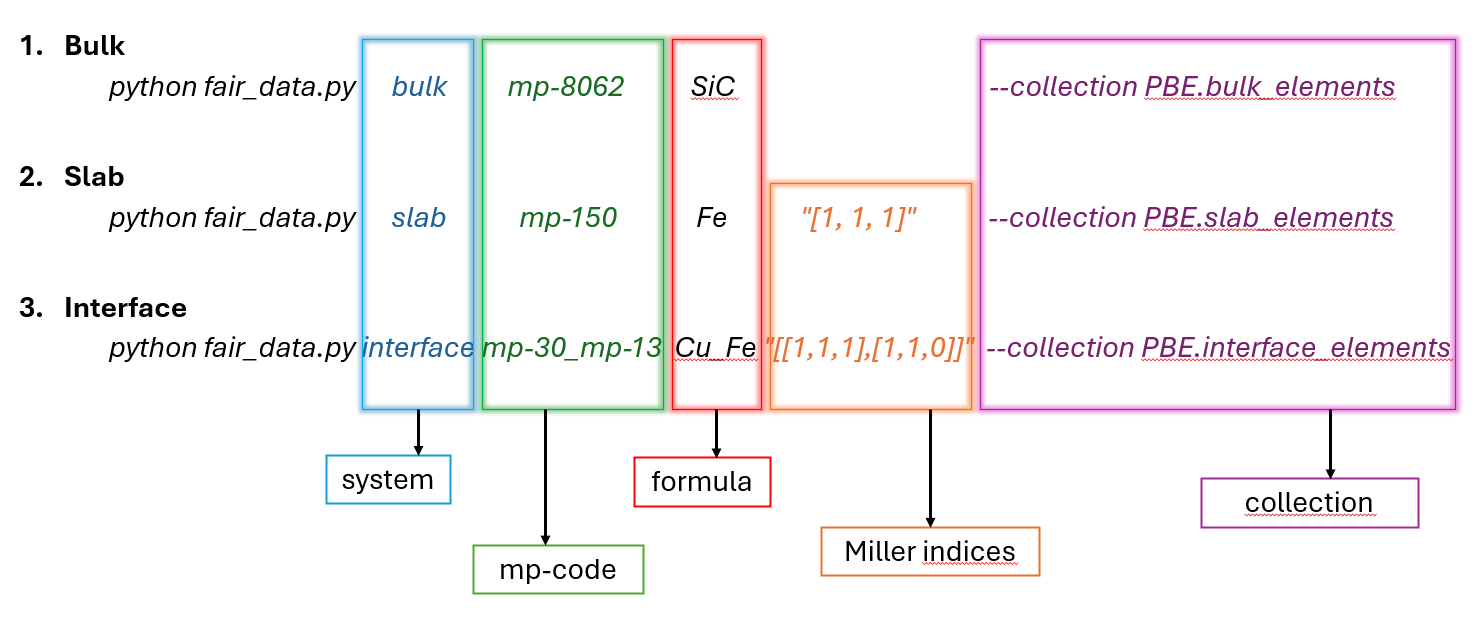}
    \caption{The figure represents the structure of the command used to run the fair data utility. System, mp-code, formula and Miller indices are mandatory arguments; collection is an optional argument instead.}
    \label{fig:usage_commands}
\end{figure*}

\texttt{fair\_data.py} is a command-line tool designed to automatically produce FAIR-compliant outputs from Tribchem calculations. It mirrors the syntax of Tribchem's own CLI, making it easy to integrate it into existing workflows. The syntax of the usage command is shown in Fig.\ref{fig:usage_commands}:
\begin{itemize}
    \item \textbf{system}: indicates the type of system among bulk/slab/interface;
    \item \textbf{mp-code}: is the code that identifies a specific system from Materials Project\cite{Materials_Project};
    \item \textbf{formula}: indicates the elements of the system;
    \item \textbf{Miller indices} of the slab for the surface and interface cases;
    \item \textbf{collection}: optional argument specifying the collection where the data should be searched. If the \texttt{collection} is not specified, the script search the document through all the collections.
    \end{itemize}

Here there are examples of usage commands for each of the different bulk/surface/interface systems:
\begin{itemize}
    \item bulk:\texttt{ python fair\_data.py bulk mp-8062 SiC --collection PBE.bulk\_elements};
    \item surface:\texttt{ python fair\_data.py slab mp-150 Fe "[1, 1, 1]" --collection PBE.slab\_elements};
    \item interface:\texttt{ python fair\_data.py interface mp-30\_mp-13 Cu\_Fe "[[1,1,1],[1,1,0]]" --collection PBE.interface\_elements};
\end{itemize}

\begin{figure*} [htbp!]
    \centering
    \includegraphics[width=1.0\linewidth]{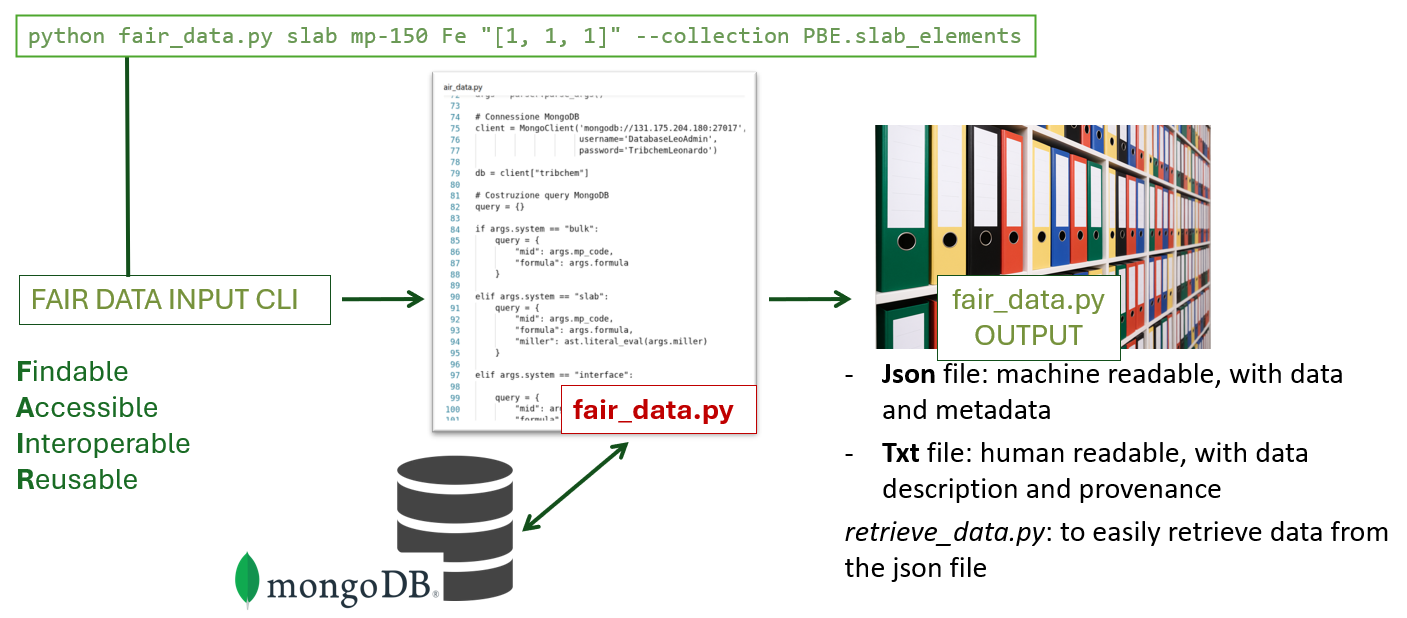}
    \caption{Schematic representation of the fair utility (for the Fe (111)-surface): the user provides input via CLI the information of the specific object, the script \texttt{fair data.py} connects to MongoDB database and outputs the json and txt files containing data and metadata relative to that system.}
    \label{fig:fair_pipeline}
\end{figure*}

The script connects to MongoDB database to retrieve the specific document and produces the following output files which ensure that data is both technically robust and easy to interpret:
\begin{itemize}
    \item a JSON file: machine-readable, containing all data and metadata;
    \item a TXT file: human-readable, detailing computational setup, system configuration, and workflow used.
\end{itemize}

The TXT file is diversified for bulk, surface and interface structures but it follows the same structure in every case:
\begin{itemize}
    \item \texttt{data description};
    \item \texttt{data provenance}, diversified into \texttt{Computational setup} and \texttt{System setup};
    \item \texttt{workflow} used.
\end{itemize}

\begin{figure*} [h!]
    \centering
    \includegraphics[width=0.9\linewidth]{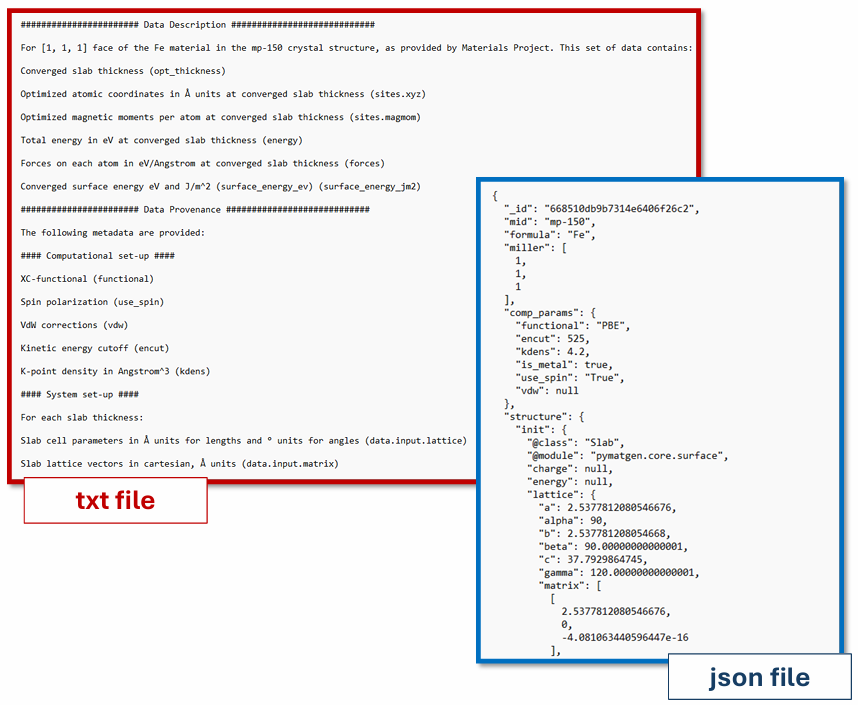}
    \caption{JSON and TXT files produced by \texttt{fair data.py} for the Fe (111)-surface in the mp-150 crystal structure.}
    \label{fig:fair_outputs}
\end{figure*}

The Fig. \ref{fig:fair_outputs} shows an example of JSON and TXT files generated for the (111)-surface of Fe having mp-code 150.

\begin{figure*} [htbp!]
    \centering
    \includegraphics[width=0.9\linewidth]{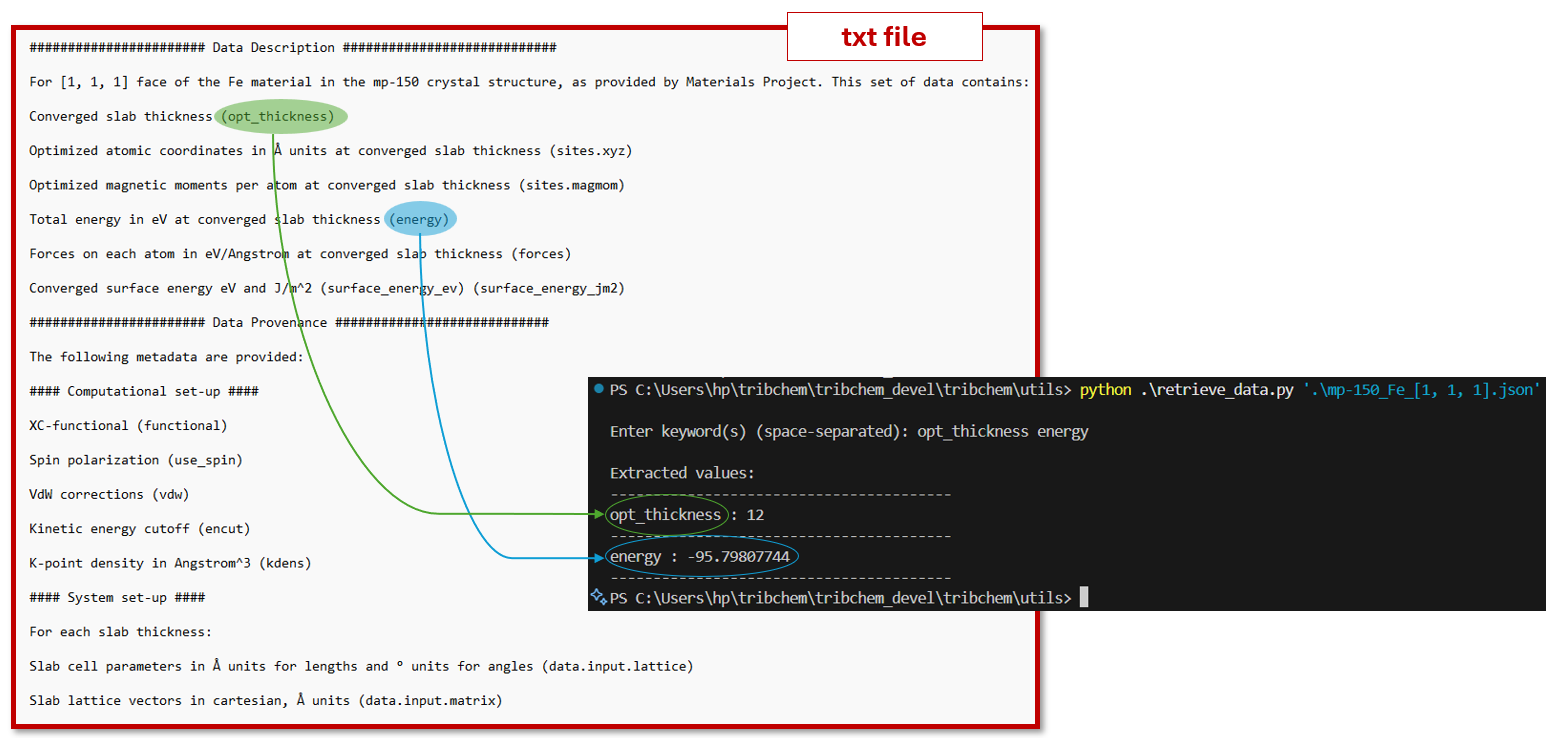}
    \caption{Usage example of \texttt{retrieve\_data.py}: the keywords are listed in a \texttt{.txt} file and passed through the terminal to retrieve the corresponding quantities.}
    \label{fig:keywords}
\end{figure*}

In the TXT file, the quantities of interest are associated to keywords, as shown in Fig.\ref{fig:keywords}, that allow the script \texttt{retrieve\_data.py}, described in the next subsection, to retrieve the specific data from the JSON file.

\subsection{\texttt{retrieve\_data.py} script}
To simplify data access by an interested human, the script \texttt{retrieve data.py} can be used to extract specific information from the JSON file. This utility facilitates efficient data analysis and reuse, aligning with FAIR standards.

\begin{figure*} [htbp!]
    \centering
    \includegraphics[width=0.9\linewidth]{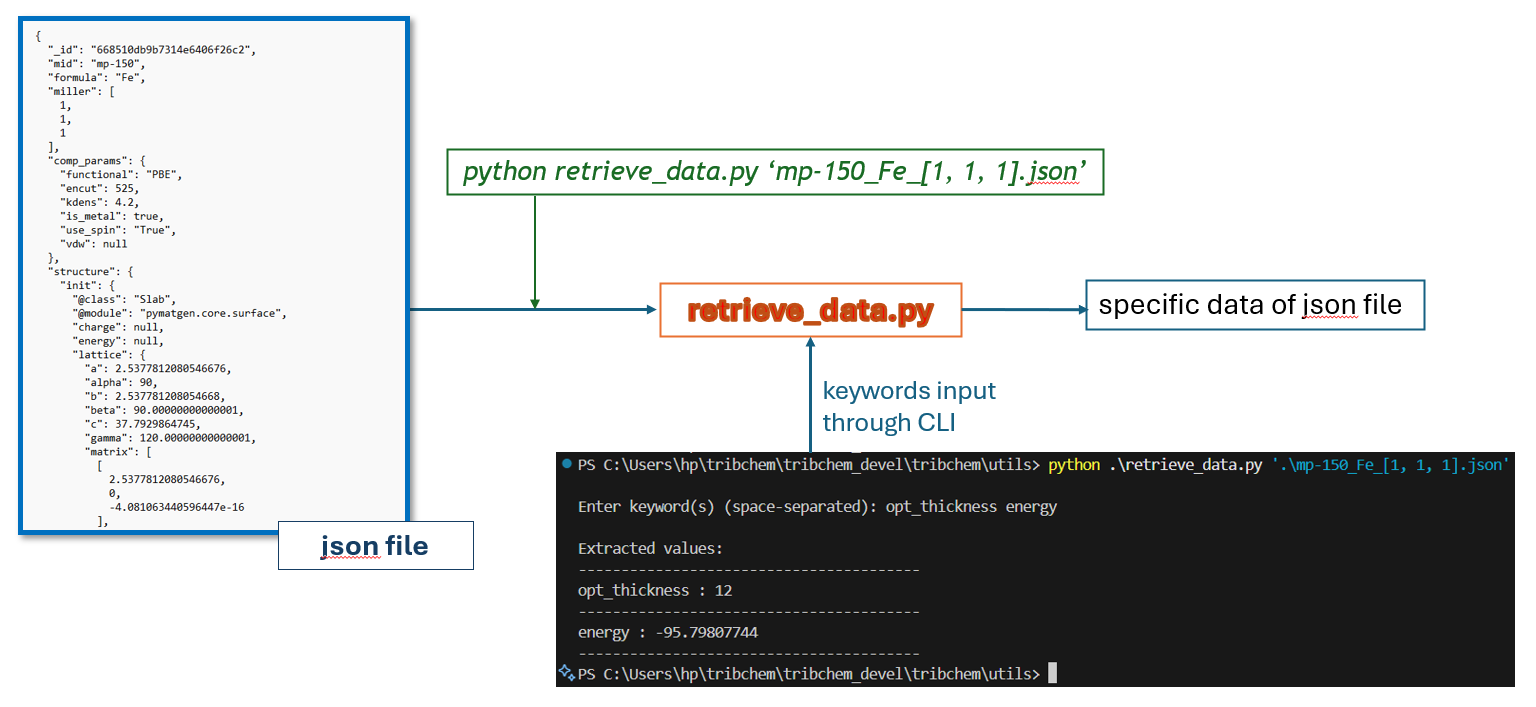}
    \caption{Schematic representation of the \texttt{retrieve\_data.py} execution pipeline, including input parsing and CLI-based keyword selection for data retrieval.}
    \label{fig:retrieve_pipeline}
\end{figure*}

The execution of \texttt{retrieve\_data.py} is shown in Fig.\ref{fig:retrieve_pipeline}. The script takes as input the JSON file specified in the usage command. The user is then prompted via the CLI to enter keywords corresponding to the quantities to be retrieved, which are subsequently printed to the terminal.

\section{Conclusions}

TribChem is a high-throughput code developed by our group for the study of solid-solid interfaces. In an automatized fashion it computes key interfacial properties (such as adhesion energy, potential energy surface (PES), perpendicular potential energy surface (PPES), charge displacement, and shear strength) storing all the results and relevant information in a high-level structured database.

The integration of FAIR principles into computational workflows represents a fundamental shift toward more transparent, reproducible, and collaborative scientific research. Through the development of the FAIR utility for TribChem, we have demonstrated a practical approach to seamlessly embedding FAIR compliance into existing computational workflows.

The \texttt{fair\_data.py} and \texttt{retrieve\_data.py} tools provide researchers with an efficient and user-friendly mean to transform their TribChem calculations into FAIR-compliant datasets. By maintaining the familiar command-line interface syntax of TribChem while automatically generating both machine-readable JSON files and human-readable TXT documentation, these utilities enable a smooth transition to FAIR practices within existing research frameworks.
The implementation addresses several critical aspects of the FAIR principles: enhanced findability through comprehensive metadata generation, improved accessibility via standardized file formats, increased interoperability through JSON-based data structures, and enhanced reusability through detailed provenance information and computational setup documentation. These improvements not only benefit individual researchers but also strengthen the broader materials science community by facilitating data sharing, comparison, and reuse across different research groups and institutions.
Furthermore, the utility's design enables seamless integration with public repositories such as Zenodo, ensuring long-term data preservation and establishing permanent digital object identifiers for computational datasets. This capability is particularly valuable for supporting reproducible research practices and enabling the development of large-scale materials databases that can drive machine learning applications and accelerate materials discovery.
As the scientific community continues to generate increasingly complex and voluminous computational datasets, tools like these will become essential infrastructure for maintaining the principles of open science.
This work proposes concrete tools that allow the high-throughput computational results of Tribchem calculations to be systematically preserved, shared, and reused, enhancing the overall efficiency of computational materials research.

\section{Acknowledgments}
These results are part of the "Advancing Solid Interface and Lubricants by First Principles Material Design (SLIDE)" project that has received funding from the European Research Council (ERC) under the European Union's Horizon 2020 research and innovation program (Grant agreement No. 865633).
The authors acknowledge funding by the European Union - NextGenerationEU (Spoke 6 - Multiscale Modelling $\&$ Engineering Applications). 




 \bibliographystyle{elsarticle-num} 
 \bibliography{cas-refs}





\end{document}